\documentclass[aps,prl,preprint,superscriptaddress,showpacs,showkeys]{revtex4}
\usepackage{graphics}
\begin{document}

\title{Extracting the exponential behaviors in the market data}

\author{Kota Watanabe}
\email[]{watanabe@smp.dis.titech.ac.jp}
\affiliation{Department of Computational Intelligence \& Systems Science, Interdisciplinary Graduate 
      School of Science \& Engineering, Tokyo Institute of Technology, 4259-G3-52 Nagatsuta-cho,  
      Midori-ku, Yokohama 226-8502}

\author{Hideki Takayasu}
\affiliation{Sony Computer Science Laboratories,Inc., 3-14-13 Higashigotanda, Shinagawa-ku, Tokyo 
     141-0022}
\author{Misako Takayasu}
\affiliation{Department of Computational Intelligence \& Systems Science, Interdisciplinary Graduate 
      School of Science \& Engineering, Tokyo Institute of Technology, 4259-G3-52 Nagatsuta-cho,  
      Midori-ku, Yokohama 226-8502}

\begin{abstract}
We introduce a mathematical criterion defining the bubbles or the crashes in financial market price fluctuations by considering exponential fitting of the given data. 
By applying this criterion we can automatically extract the periods in which bubbles and crashes are identified. 
From stock market data of so-called the Internet bubbles it is found that the characteristic length of bubble period is about 100 days. 
\end{abstract}


\maketitle

\section{Introduction}
The analysis of bubbles or crashes in financial markets is hot study in econophysics \cite{kaizoji,mizuno1}. 
These phenomena are often big social problems like the cases of the Black Monday or the Internet bubble.
However, the definition of the bubbles or the crashes is not clear so far, namely, there is no mathematical definition or criterion formula of these phenomena. \\

In order to tackle this problem we pay attention to an empirical fact that an exponential curve fits better to bubble or crash data than the popular linear trend lines. 
By mathematically describing the exponential fitting we try to propose a mathematical definition of bubbles and crashes, so that we can automatically specify the period of a bubble and crush. \\

In this paper we analyze the high-frequency NASDAQ data focusing on the Internet bubble or the dot-com bubble appeared at the end of 20th century. 
We calculate an average of prices every thirty seconds in the tick data. Regular trading time in NASDAQ is from 9;30 to 16;00, so the number of data points in a day is 780.
\section{Extraction of the exponential behaviors}
We introduce the following formula for extracting the exponential behaviors in the financial time series. 
\begin{equation}
P(t)-P_{0}(i;T_{i})=\omega_{1}(i;T_{i})\{P(t-1)-P_{0}(i;T_{i})\}+F(t)
\end{equation}
This formula has an autoregressive form where the current state is given by the past states.
In this formula, $P(t)$ is a price at time t, $\omega_{1}(i;T_{i})$ is the parameter characterizing the 
exponential behaviors in the i-th period of length $T_{i}$,  If $\omega_{1}(i;T_{i})$ is larger than 1.0,  
the time series are either exponentially increasing or decreasing, then $P_{0}(i;T_{i})$ gives the base line of these exponential divergence.
If $\omega_{1}(i;T_{i})$ is less or equal to 1.0, it means that there is no bubble-like trend or the time series is convergent, then $P_{0}(i;T_{i})$ shows an asymptotic line.
$F(t)$ is residual noise term. The parameters $\omega_{1}(i;T_{i})$ and $P_{0}(i;T_{i})$ can be determined uniquely under the condition that minimizes the errors, which is the sum of squares of $F(t)$.
\section{Estimation of the optimal period}
For applying Eq.(1) to the time series, we need to estimate the length of the period $T_{i}$  
that can be fitted by an exponential function. We introduce a minimum period of $T_{i}$ by using 
the following well-known auto-regressive (N) model for the price difference time series.
\begin{eqnarray}
\Delta P(t) &=& \displaystyle \sum_{j=1}^{j=N-1}b_{j} \Delta P(t-j)+f(t) \\
\Delta P(t) &=& P(t)-P(t-1)  
\end{eqnarray}
Here, {$b_{j}$} give the AR parameters that make the residue, $f(t)$, almost an independent random 
noise \cite{ohnishi1,ohnishi2,mizuno2}. In this equation we tune the parameters of AR so that the standard deviation of $f(t)$ 
as similar as possible to the real stock price data such as Yahoo! Inc.(ticker symbol YHOO).
Now we define the time scale of $T_{i}$ by the minimum time scale that satisfies the condition where $\omega_{1}(i;T_{i})$ is always less or equal to $1.0$ 
when the time series are created by the Eqs. 
(2) and (3) with N=5. Actually, by changing $T_{i}$ from 1 day to 100 days, the frequency of finding $\omega_{1}(i;T_{i})$ larger than 1.0 decreases. 
When we set $T_{i}$ to be longer than 100 days, we cannot observe $\omega_{1}(i;T_{i})$ to take a value larger than 1.0 in practical sense. 
Therefore, we fix the optimal time scale $T_{i}$ for observing the exponential behaviors 
to be 100 days. On the basis of this AR analysis, if we observe $\omega_{1}(i;T_{i})$ larger than 1.0 in the time range of 100 days in real data, 
we can say that the real time series fluctuation of that range is statistically different from the AR model which implies that a non-stationary description is needed.
\begin{figure}[htbp]
\includegraphics[9.0cm,5.0cm]{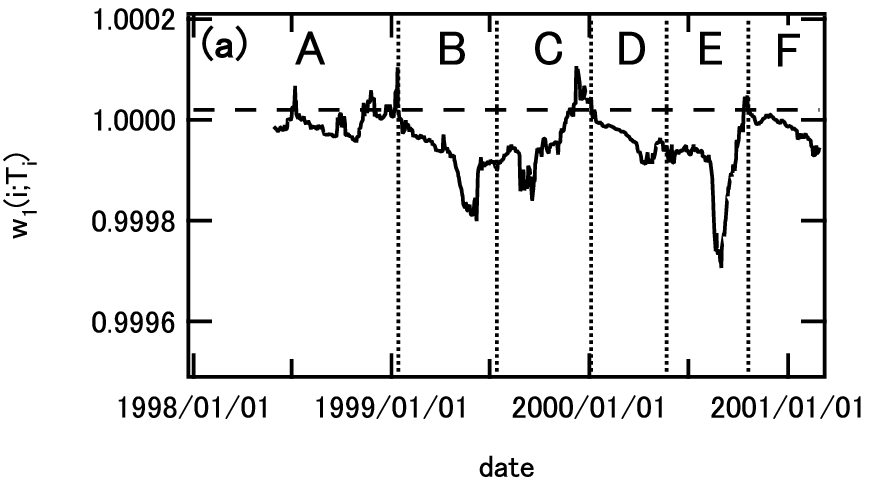}%
\includegraphics[9.0cm,5.0cm]{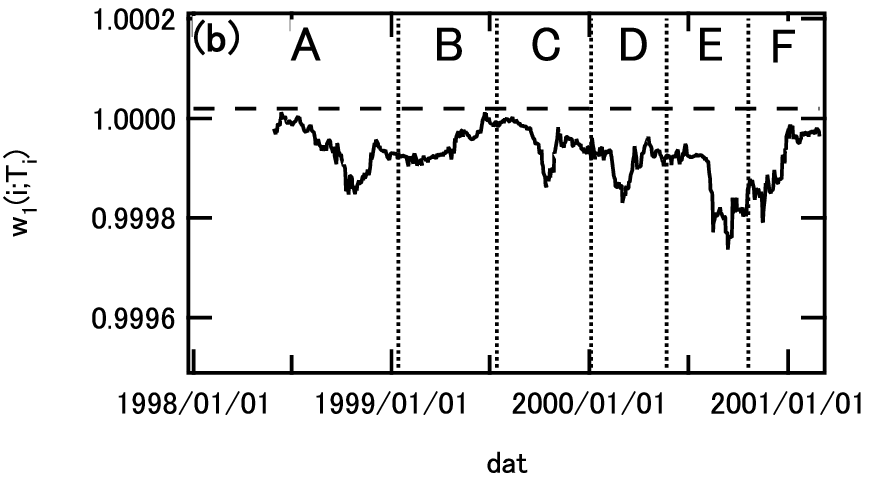}%
\caption{The time series of $\omega_{1}(i;T_{i})$ with  $T_{i}$ =100 days. (a) The case of YHOO (b) Random walk based on AR(5). }
\label{fig2}
\end{figure} 
\section{Assignment of bubble, crash and convergence}
We now assign each time step (one time step is thirty seconds) either exponential or convergent.
If the observing box of 100 days is judged as exponentially diverging, that means $\omega_{1}(i;T_{i})$ is larger than 1.0 in the box, 
we assign all time steps in the box as exponential. Then, we shift the box by one time step, and calculate $\omega_{1}(i;T_{i})$ for the new box. If the value of $\omega_{1}(i;T_{i})$ is less than 1.0, then only the latest time step is assigned as convergent.
Repeating this process to cover all the data we can separate the exponential periods and convergent periods,
see Fig.2. At this stage the length of each period takes any value independent of the observing time scale $T_{i}$. 
Note that we can detect the start of slow exponential behavior before the extreme price fluctuations as found in Fig.3.  \\
\begin{figure}[htbp]
\includegraphics[14cm,8cm]{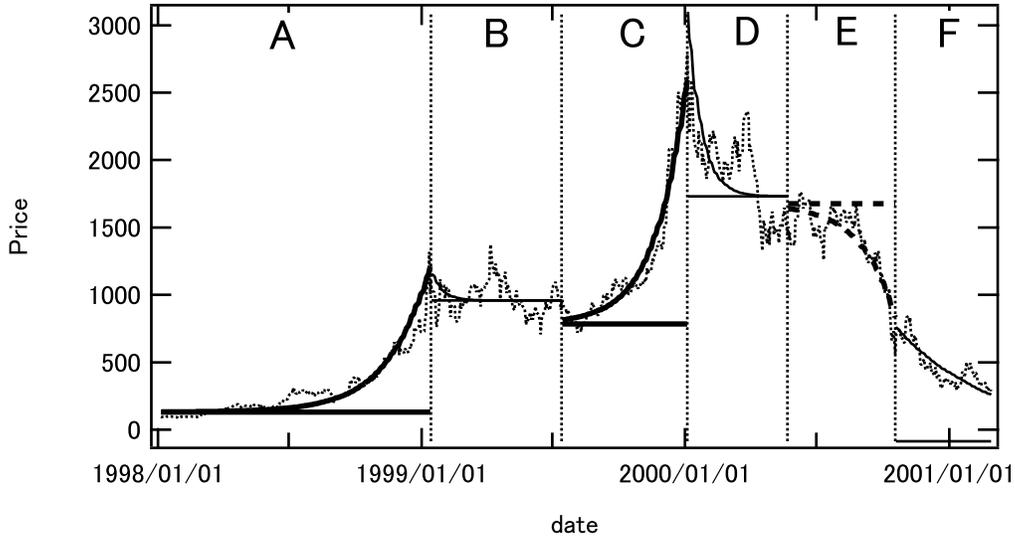}%
\caption{The exponential trend curves and $P_{0}(i;T_{i})$ in each period. Time series of YHOO (dotted line). The bubble periods (heavy line) are A and C. The crash period (heavy dashed line) is E. The convergent period (line) are B, D and F.}
\label{fig2}
\end{figure} 
\begin{figure}[htbp]
\includegraphics[14cm,8cm]{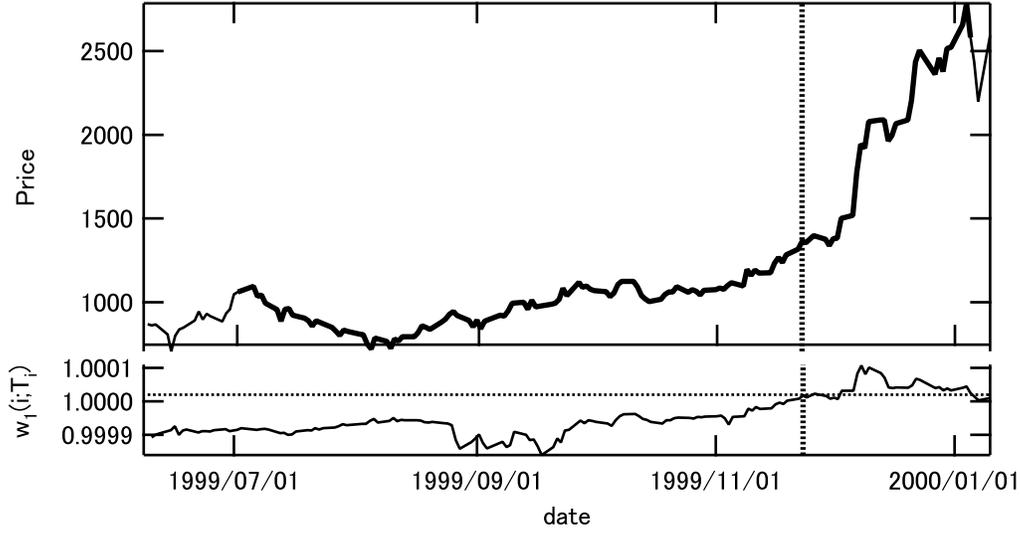}%
\caption{The time series of a price and $\omega_{1}(i;T_{i})$ in the     
    scale from the end of the period B to the beginning of the period D. The vertical dotted line   
    shows that the first point of detecting the bubble.}
\label{fig1}
\end{figure}

Next, we calculate the parameters $\omega_{1}(i;T_{i})$ and $P_{0}(i;T_{i})$ to each period by applying Eq.(1) again. 
Then, we can draw an exponential trend curve for each exponential period as shown in      
Fig.2 by using the following equation.
\begin{equation}
P_{trend}(t)=\omega_{1}(i;T_{i})P_{trend}(t-1)+(1-\omega_{1}(i;T_{i}))P_{0}(i;T_{i})
\end{equation}
In this equation $P_{trend}(t)$ is the exponential trend price at time t. When the exponential trend is diverging upward we call it as "bubble", when it is diverging downward it is called as "crash". 
When the trend curve is converging we call it as "convergence". 
The starting value of this price in each period is derived from theoretical calculation. From Fig.2 we find that the exponential behaviors are continuing more than 250 days in the period A. 
A critical case may be the period D. As this period starts with a sharp drop, it may intuitively look a kind of "crash", however, according to our analysis the trend line shows a convergence to a lower price.\\

In Fig.4 we compare the error estimation between our exponential approximation and the usual linear approximation in each period. The errors $E(i)$ are calculated by the following equation.
\begin{equation}
E(i)=\sqrt{<(P(t)-P_{trend}(t))^{2}>}
\end{equation}
The linear approximation is determined by the least-square-method. We can find that errors 
become smaller for exponential approximation compared to the linear approximation in cases of
the bubbles or crashes (the periods A,C and E).  \\

The residual error $F(t)$ in Eq. (1) is not perfectly uncorrelated because just 
extracting the exponential trends in big scale can not extract all local trends. In order to separate pure noises we apply the 
Yule-Walker formula for the time series of $F(t)$
\begin{equation}
F(t)=\displaystyle \sum_{j=1}^{j=N-1}a_{j}(i;T_{i})F(t-j)+N(t)
\end{equation}
Then, Eqs.(1) and (7) make the following equation.
\begin{equation}
P(t)=\displaystyle \sum_{j=1}^{N}\Omega_{j}(i;T_{i})P(t-j)+(1-\sum_{j=1}^{N}\Omega_{j}(i;T_{i}))P_{0}(i;T_{i})+N(t)
\end{equation}
In this equation $\Omega_{j}(i;T_{i})$ is written by the following forms.
\begin{eqnarray}
\Omega_{1}(i;T_{i}) &=& a_{1}(i;T_{i})+\omega_{1}(i;T_{i}) \\ 
\Omega_{j}(i;T_{i}) &=& a_{j}(i;T_{i})-a_{j-1}(i;T_{i})\omega_{1}(i;T_{i})~~~~~(2 \le j \le N-1 )  \\ 
\Omega_{N}(i;T_{i}) &=& -a_{N-1}(i;T_{i})\omega_{1}(i;T_{i})
\end{eqnarray}
Here the noise term $N(t)$ is confirmed to be nearly perfectly uncorrelated.
\begin{figure}[htbp]
\includegraphics[14cm,8cm]{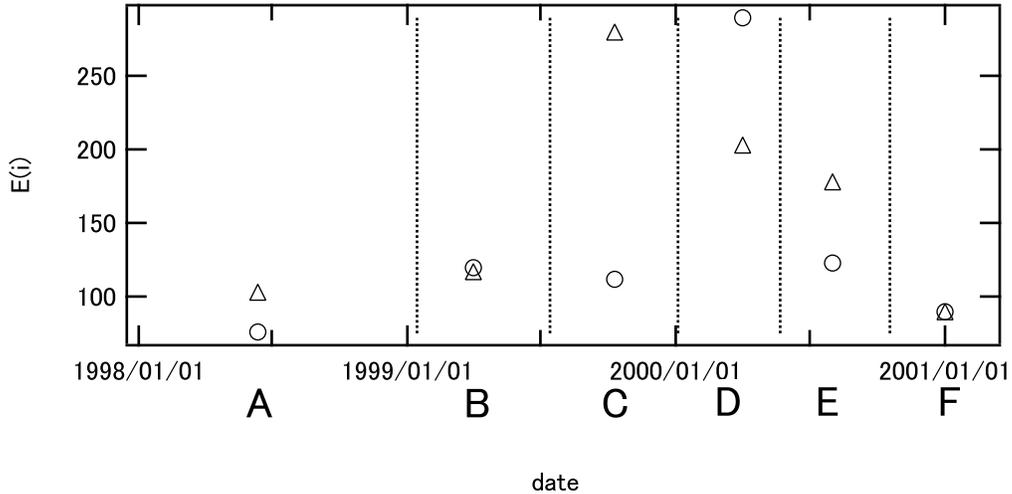}%
\caption{Comparison of the errors $E(i)$ in each periods. Exponential approximations (circles) and linear approximations (triangles). }
\label{fig3}
\end{figure} 
\section{Conclusion}
In this paper we have mathematically defined bubbles and crashes by the exponential behaviors.   
As we have discussed in proceeding section this method can be used for prediction of  large price changes in macroscopic scale. 
However, the predictive information in our method is not enough, for example, we can not tell when a bubble will stop. 
As a future work we are now combining the potential analysis method with this bubble detection technique to clarify the mechanism of bubbles and crashes \cite{takayasu1,takayasu2,takayasu3}.
\section{Acknowledgement}
\begin{acknowledgments}
This work is partly supported by Japan Society for the Promotion of Science, Grant-in-Aid for Scientific Research $ \sharp $ 16540346 (M.T.).
\end{acknowledgments}

\bibliography{REFERENCE}

\end{document}